\documentclass[preprintnumbers,eqsecnum,floatfix]{revtex4}
\usepackage{amsfonts}
\usepackage{amsmath}
\usepackage{amssymb}
\usepackage{graphicx}
\usepackage{setspace}

\begin{document}

\title{Two Gamma Quarkonium and Positronium Decays with Two-Body Dirac
Equations of Constraint Dynamics}
\author{Horace W. Crater}
\affiliation{The University of Tennessee Space Institute, Tullahoma, TN 37388\thanks{%
hcrater@utsi.edu}}
\author{Cheuk-Yin Wong}
\affiliation{Department of Physics, University of Tennessee, Knoxville, TN 37996}

\begin{abstract}
Two-Body Dirac equations of constraint dynamics provide a covariant
framework to investigate the problem of highly relativistic quarks in meson
bound states. \ This formalism eliminates automatically the problems of
relative time and energy, leading to a covariant three dimensional formalism
with the same number of degrees of freedom as appears in the corresponding
nonrelativistic problem. It provides bound state wave equations with the
simplicity of the nonrelativistic Schr\"{o}dinger equation. Unlike other
three-dimensional truncations of the Bethe-Salpeter equation, this covariant
formalism has been thoroughly tested in nonperturbatives contexts in QED,
QCD, and nucleon-nucleon scattering.\ Here we continue the important studies
of this formalism by extending a method developed earlier for positronium
decay into two photons to tests on the sixteen component quarkonium wave
function solutions obtained in meson spectroscopy. We examine positronium
decay and then the two-gamma quarkonium decays of $\eta_{c},\eta_{c}^{%
\prime},\chi _{c0},\chi_{c2},$ and $\pi^{0}.$ \ The results for the $\pi^{0}$%
, although off the experimental rate by 13\%, is much closer than the usual
expectations from a potential model.\ 
\end{abstract}

\eid{}
\startpage{1}
\endpage{1}
\maketitle

\vspace*{-0.8cm}

\section{Constraint Two-Body Dirac Equations for QED and QCD}

\vspace*{-0.2cm}

The Bethe-Salpeter equation (BSE) can be derived from QFT but its
implementation invariably involves a non-unique choice from a very large
class of three dimensional truncations, almost all of which work well in the
perturbative context. We describe one such approach, the Two-Body Dirac
Equations (TBDE), based on Dirac's \ constraint dynamics that have been
successfully applied to two-body bound state problems in QED \cite{bckr} and
QCD \cite{cr04} and to two-body nucleon-nucleon scattering \cite{liu}. We
will describe various nonperturbative tests we have applied to distinguish
it from the others \cite{cr04}. We then describe its most recent test in the
treatment of two-photon decays of positronium and quarkonium \cite{cr06}.

The TBDE can be derived from the BSE but they had their origins in classical
relativistic mechanics \cite{szcr}. For spin-zero particles one constructs
two generalized mass shell constraints $\mathcal{H}_{i}\equiv
p_{i}^{2}+m_{i} ^{2}+\Phi_{i}~\mathcal{\approx}~0;~i=1,2$ and guarantees
their compatibility by requiring that the potentials satisfy a relativistic
third-law condition $\Phi_{1}=\Phi_{2}=\Phi(x_{\perp},p_{1},p_{2})\equiv%
\Phi_{w}~$and depend only on the interparticle separation component
perpendicular to the total momentum, 
\begin{equation}
x_{12\perp}^{\mu}=(\eta^{\mu\nu}+\hat{P}^{\mu}\hat{P}^{\nu})(x_{1}-x_{2}
)_{\nu};~P^{\mu}=p_{1}^{\mu}+p_{2}^{\mu}~~;\ ~\hat{P}^{\mu}\equiv~P^{\mu }/%
\sqrt{-P^{2}}~;~~x_{12\perp}\cdot~\hat{P}=0.
\end{equation}
Thus, interactions depend on the invariant $\sqrt{x_{\perp}^{2}} =\sqrt{%
x^{2}+(x\cdot\hat{P})^{2}} \equiv r$ which reduces to the spatial separation 
$\sqrt{\mathbf{r}^{2}}$ in the c.m frame, where the relative time cancels
out covariantly. Calculating the difference of these compatible constraints, 
$\mathcal{H} _{1}-\mathcal{H}_{2}$$=$$-2P\cdot p~\mathcal{\approx}~0$, gives
a covariant elimination of the c.m. relative energy, complimentary to the
relative time restriction. The relative momentum $p^{\mu}$$=$ $%
(\varepsilon_{2}p_{1}^{\mu }-\varepsilon_{1}p_{2}^{\mu})/w$ is given in
terms of the total c.m. energy $w=\varepsilon_{1}+\varepsilon_{2}~$and the
c.m. constituent energies $\varepsilon_{1}-%
\varepsilon_{2}=(m_{1}^{2}-m_{2}^{2})/w$. It is canonically conjugate to $%
x_{\perp}$ in a covariant three-dimensional way, $\{x_{\perp}^{\mu},p^{\nu}%
\}=\eta_{\perp}^{\mu\nu}.$ The other independent combination of the two
constraints yields a dynamical equation$~(\varepsilon _{2}\mathcal{H}%
_{1}+\varepsilon_{1}\mathcal{H}_{2})/w=p_{\perp}^{2}+\Phi _{w}-b^{2}(w)~%
\mathcal{\approx}~0$ \ whose quantization provides a covariant Schr\"{o}%
dinger-like equation ($p_{\perp}^{2}+\Phi_{w})\psi=b^{2}(w)\psi~$with the
triangle function 
\begin{equation}
b^{2}(w)=(w^{4}-2w^{2}(m_{1}^{2}+m_{2}^{2})+(m_{1}^{2}-m_{2}^{2})^{2}
)/4w^{2}=\varepsilon_{w}^{2}-m_{w}^{2}
\end{equation}
playing the role the eigenvalue. Its appearance signals exact two-body
kinematics with effective particle motion displaying an Einstein relation
between energy $\varepsilon_{w}=(w^{2}-m_{1}^{2}-m_{2}^{2})/2w$, mass $%
m_{w}=m_{1}m_{2}/w$, and momentum.

For quantum mechanical spinning particles one has two Dirac equations (here
given for minimal vector and scalar interactions) instead of two generalized
Klein-Gordon-like mass shell constraints. 
\begin{equation}
\mathcal{S}_{i}\psi\equiv\gamma_{5i}(\gamma_{i}\cdot(p_{i}-\tilde{A}
_{i})+m_{i}+\tilde{S}_{i})\psi=0;~~~i=1,2.
\end{equation}
As in the spinless case they are compatible, $[\mathcal{S}_{1},\mathcal{S}
_{2}]\psi=0$, provided that supersymmetry is added to the conditions that
apply in the two body spinless case. One finds that the constituent scalar $%
\tilde{S}_{i}=\tilde{S}_{i}(S(r),A(r),p_{\perp},w,\gamma_{1},\gamma_{2})~$%
and vector $\tilde{A}_{i}^{\mu}=\tilde{A}_{i}^{\mu}(A(r),p_{\perp},w,\gamma
_{1},\gamma_{2})~$potentials are spin-dependent with the dynamics arising
from single invariant functions, one for each type of interaction.

One of the advantages this covariant three dimensional equation has over the
Bethe Salpeter equation is its simplicity. We obtain an invariant three
dimensional Schr\"{o}dinger-like equation 
\begin{equation}
{\Large \{} p^{2}+\Phi_{w}(\sigma_{1},\sigma_{2},p_{\perp},A(r),S(r)) 
{\Large \}} \psi=b^{2}(w)\psi  \label{inv}
\end{equation}
in terms of a sixteen component wave function $\psi=[\psi_{1},\psi_{2},
\psi_{3},\psi_{4}]$. Explicitly one finds that the equation has the
following form exhibiting coupling between the upper-upper and lower-lower
components in addition to the usual spin-dependent interactions symbolized by%
$~\Phi (\sigma_{1},\sigma_{2},p_{\perp},A(r),S(r))$, 
\begin{eqnarray}
\{p^{2}+2m_{w}S+S^{2}+2\mathcal{\varepsilon}_{w}A-A^{2}+\Phi_{11}\}\psi
_{1}+\Phi_{14}\psi_{4}-b^{2}(w)\psi=0.
\end{eqnarray}
This equation and the one for the lower-lower component are quantum
mechanically well defined, allowing nonpertubative numerical solutions. A
second advantage is that in the case of lowest order QED where $%
A(r)=-\alpha/r$ they have an \nolinebreak analytic Sommerfeld-like solution
for singlet positronium with spectral results that agree with standard
perturbative results with the same spectral agreement holding numerically
with triplet states \cite{bckr}. This implies that they are less likely to
produce spurious physics when applied to QCD as may occur in formalisms that
do not have this type of nonperturbative QED agreement. \ Without successful
nonperturbative QED tests how can a candidate two-body formalism be trusted
for QCD spectral results?

For QCD bound states we use a covariant version of the Adler-Piran static
quark potential \cite{adlr} $V_{AP}(r)=\Lambda(U(\Lambda r)+U_{0} )\ (=A+S)~$%
obtained from an effective non-linear Maxwell equation embodying QCD. It
analytically displays both asymptotic freedom $\Lambda U(\Lambda
r<<1)\sim1/r\ln\Lambda r~$and linear quark confinement, $\Lambda U(\Lambda
r>2)\sim\Lambda^{2}r$. We obtain a covariant reinterpretation of their
static model by replacing $r$ with $x_{\perp}$ and apportioning the
potential between our vector and scalar invariants $A$ and $S$ so that at
short distance it is vector and long distance it is scalar \cite{cr04}.~Once
the underlying vector and scalar invariants and quark masses are fixed so
are all the spectral predictions from Eq.(\ref{inv}).

We obtain very good results for the entire meson spectrum from the light
pion to the heavy upsilon states. The quality of our fit is about the same
obtained by Godfry and Isgur \cite{isg}, but with just 2 invariant functions
instead of their 6 \cite{cr04}. The relativistic coupling structure of our
equations are equally important for positronium and $\pi-\rho$ system. If we
ignore the coupling to the lower-lower components $\psi_{4}$, we obtain poor
hyperfine splitting results for the positronium system. Likewise we would
obtain poor results for the hyperfine $\pi$$-$$\rho$ splitting if we ignored
that coupling ($m_{\pi}\sim 850$ MeV; $m_{\rho}\sim1060 $ MeV) instead of
the fully coupled results $m_{\pi}\sim 144$ MeV; $m_{\rho}$$~$$\sim792$ MeV 
\cite{cr04}.

The nonperturbative structures in our equations provide for chiral symmetry
in that the pion (although not its excited states or $\rho$) behaves like a
Goldstone boson. With the coupling structure, the pion mass tends to zero as
the quark mass tends to zero, but without the coupling to lower-lower
component this does not occur. However, the interaction structure of the
TBDE when restricted to vector and scalar interactions does not give the
functional dependence of $m_{\pi}$ on the quark mass of $m_{\pi}^{2}\sim
m_{q}$ behavior dictated from the nonconserved axial current generator of
chiral symmetry. \ 

\vspace*{-0.2cm}

\section{Two-Photon Decays from Two-Body Dirac Equation Wave Functions}

\vspace{-0.2cm}

We take advantage of this close relation between the pion and positronium to
treat two photon meson and positronium decays equally. The on-shell Feynman
amplitude for electron-positron annihilation in the singlet state is 
\begin{equation}
M_{\alpha\beta}=\frac{e^{2}}{(2\pi)^{3}w\sqrt{2}}\{\bar{v}^{(s_{+})} (p_{+})%
{\large [}\gamma\cdot\epsilon^{(\alpha_{1})}\frac{m-i\gamma\cdot
(p_{-}-k_{1})}{(p_{-}-k_{1})^{2}+m^{2}-i0}\gamma\cdot\epsilon^{(\alpha_{2}
)}+(1\Leftrightarrow2)]u^{(s_{-})}(p_{-})-(s_{+}\Leftrightarrow s_{-})\}
\end{equation}

The two-step change from above amplitude to that for positronium
annihilation in which constituents are off shell is indicated below 
\begin{align}
M_{\alpha\beta} & \rightarrow\int d^{3}p\tilde{\psi}(\mathbf{p)}
M_{\alpha\beta}\equiv\mathcal{M}_{^{1}S_{0}\rightarrow2\gamma}\equiv\int
d^{3}p\tilde{\psi}(\mathbf{p)}\bar{v}(-\mathbf{p)\Gamma(p,k)}u(\mathbf{p} ) 
\notag \\
& =\int d^{3}p\tilde{\psi}(\mathbf{p)}Tr\mathbf{\Gamma(p,k)}u(\mathbf{p} )%
\bar{v}(-\mathbf{p)}\rightarrow\int d^{3}pTr\mathbf{\Gamma(p,k)} \Psi\mathbf{%
(\mathbf{p)}}  \label{gm}
\end{align}
in which in the final step the 4x4 matrix wave function $\Psi~$replaces the
outer product of the free $u,\bar{v}$ spinors.

In Schr\"{ }odinger-like form the TBDE for singlet positronium becomes 
\begin{equation}
(-\nabla ^{2}-2\varepsilon _{w}{\alpha }/r-\alpha ^{2}/r^{2})\psi (\mathbf{r}%
)=b^{2}(w)\psi (\mathbf{r}),
\end{equation}%
with an exact solution wave function mildly singular at the origin. However,
unlike standard approaches with rates${\LARGE \ }$proportional to $%
\left\vert \psi (0)\right\vert ^{2}~$(which would give divergent results)
the configuration space form of Eq.(\ref{gm}) below smears the annihilation
amplitude over a \ Compton wave length so that singularities are rendered
harmless: 
\begin{equation}
\mathcal{M}_{^{1}S_{0}\rightarrow 2\gamma }=e^{2}\sqrt{\pi /2}\int
d^{3}r\exp (-i\mathbf{k\cdot r)}Tr\{\psi (\mathbf{r)[\gamma \cdot \hat{%
\varepsilon}}_{1}(m+i\mathbf{\gamma \cdot \nabla )\gamma \cdot \hat{%
\varepsilon}}_{2}+\mathbf{\gamma \cdot \hat{\varepsilon}}_{2}(m+i\mathbf{%
\gamma \cdot \nabla )\gamma \cdot \hat{\varepsilon}}_{1}]\exp (-mr)/r\}.
\label{amp}
\end{equation}%
This gives rise to standard decay width formulae for positronium: 
\begin{align}
\Gamma (^{1}S_{0}& \rightarrow 2\gamma )=m\alpha ^{5}/2,  \notag \\
\Gamma (^{3}P_{0}& \rightarrow 2\gamma )=3m\alpha ^{7}/256,~~~\Gamma
(^{3}P_{2}\rightarrow 2\gamma )=m\alpha ^{7}/320,  \notag \\
\Gamma (^{3}P_{0}& \rightarrow 2\gamma )/\Gamma (^{3}P_{2}\rightarrow
2\gamma )=15/4.
\end{align}%
We recast the equations $\mathcal{S}_{i}\psi =0$ in terms of the mass and
energy potentials 
\begin{align}
M_{i}& =m_{i}\ \cosh L(S,A)\ +m_{i}\sinh L(S,A)  \notag \\
E_{i}& =\epsilon _{i}\ \cosh \mathcal{G}(A)\ -\epsilon _{i}\sinh \mathcal{G}%
(A);~G=\exp \mathcal{G},
\end{align}%
for which 
\begin{equation}
\mathcal{S}_{i}=G\beta _{i}\Sigma _{i}\cdot p+E_{i}\beta _{i}\gamma
_{5i}+M_{i}\gamma _{5i}-(i/2)G\Sigma _{j}\cdot \partial (\mathcal{G}\beta
_{i}+L\beta _{2})\gamma _{51}\gamma _{52};~~i,j=1,2.
\end{equation}%
For the pion and other 2-$\gamma $ meson decays we need to transform from 16
component wave function solutions of TBDE to their 4x4 matrix forms. The
computation is simplified by first obtaining decoupled Schr\"{ }odinger like
form of the TBDE through use of the plus/minus \ combinations $\phi _{\pm
}=\psi _{1}\pm \psi _{4};\ \ \chi _{\pm }=\psi _{2}\pm \psi _{3}~$of the
original TBDE wave functions. A further spin-dependent scale transformation
to $\psi _{\pm },\eta _{\pm }$ by the c.m. transformation 
\begin{equation}
\phi _{\pm }=\exp (F+K\boldsymbol{\sigma}_{1}\mathbf{\cdot \hat{r}}%
\boldsymbol{\sigma}_{2}\mathbf{\cdot \hat{r}})\psi _{\pm };~\chi _{\pm
}=\exp (F+K\boldsymbol{\sigma}_{1}\mathbf{\cdot \hat{r}}\boldsymbol{\sigma}%
_{2}\mathbf{\cdot \hat{r}})\eta _{\pm }
\end{equation}%
gives a still simpler decoupled form (with no momentum-dependent
interactions), 
\begin{align}
& \{\mathbf{p}^{2}+2m_{w}S(\left\vert \mathbf{r}\right\vert
)+S^{2}(\left\vert \mathbf{r}\right\vert )+2\varepsilon _{w}A(\left\vert 
\mathbf{r}\right\vert )-A^{2}(\left\vert \mathbf{r}\right\vert )+\Phi
_{D}(S,A,\nabla S,\nabla A,\nabla ^{2}S,\nabla ^{2}A)  \notag \\
& \mathbf{+L\cdot (}\boldsymbol{\sigma}_{1}\mathbf{+}\boldsymbol{\sigma}_{2}%
\mathbf{)}\Phi _{SO}(..)+\mathbf{L\cdot (}\boldsymbol{\sigma}_{1}\mathbf{-}%
\boldsymbol{\sigma}_{2}\mathbf{)}\Phi _{SOD}(..)+i\mathbf{L\cdot }%
\boldsymbol{\sigma}_{1}\mathbf{\times }\boldsymbol{\sigma}_{2}\Phi _{SCO}(..)
\notag \\
& +\boldsymbol{\sigma}_{1}\mathbf{\cdot }\boldsymbol{\sigma}_{2}\Phi
_{SS}(..)+\boldsymbol{\sigma}_{1}\mathbf{\cdot \hat{r}}\boldsymbol{\sigma}%
_{2}\mathbf{\cdot \hat{r}}\Phi _{T}(..)~+\boldsymbol{\sigma}_{1}\mathbf{%
\cdot \hat{r}}\boldsymbol{\sigma}_{2}\mathbf{\cdot \hat{r}L\cdot (}%
\boldsymbol{\sigma}_{1}\mathbf{+}\boldsymbol{\sigma}_{2}\mathbf{)}\Phi
_{SOT}(..)\}\psi _{+}=b^{2}(w)\psi _{+}.  \notag
\end{align}

This four-component Schr\"{o}dinger-like form has familiar spin-dependent
interactions depending on the underlying invariants $S$ ~and $A$ that govern
the scalar and vector interactions. Each of the four-component spinor wave
functions is replaced by a 2x2 matrix wave functions 
\begin{equation}
\mathcal{\phi }_{\pm }\equiv \phi _{\pm 0}1+\boldsymbol{\phi }_{\pm }\cdot %
\boldsymbol{\sigma}\mathbf{,~}\mathcal{\chi }_{\pm }\equiv \chi _{\pm 0}1+%
\boldsymbol{\chi}_{\pm }\cdot \boldsymbol{\sigma}
\end{equation}%
given in terms of singlet (scalar) ($\phi _{0\pm },\chi _{0\pm },\psi _{0\pm
},\eta _{0\pm })$ and triplet (vector) ($\boldsymbol{\phi }_{\pm }\mathbf{,}%
\boldsymbol{\chi}_{\pm }\mathbf{,}\boldsymbol{\psi }_{\pm }\mathbf{,}%
\boldsymbol{\eta}_{\pm })~$wave functions.

The decay amplitude requires the full 4x4 matrix wave function\ \ \ \ 
\begin{align}
\psi (\mathbf{r)}& \mathbf{=}\exp (F)[\cosh K\Psi (\mathbf{r)}-\sinh K%
\boldsymbol{\Sigma}\cdot \mathbf{\hat{r}}\Psi \mathbf{(\mathbf{r)}}%
\boldsymbol{\Sigma}\cdot \mathbf{\hat{r}]}  \notag \\
\Psi (\mathbf{r)}& \mathcal{=}(\mathcal{\psi }_{+}q_{1}+\mathcal{\psi }%
_{-}iq_{2}+\mathcal{\eta }_{+}q_{0}\mathbf{+}\mathcal{\eta }_{-}q_{3})/(2%
\sqrt{2});~q_{i}q_{j}=\delta _{ij}q_{0}+i\varepsilon _{ijk}q_{k}  \label{frm}
\end{align}%
Once we solve for four component spinor $\psi _{+}$ we can use the two-body
Dirac equation to solve for the remaining 12 components, $\psi _{0-},%
\boldsymbol{\psi }_{-},\eta _{0+,-}$ $\boldsymbol{\eta }_{+,-}$, with the
final connection to the numerical meson wave functions from writing $\Psi (%
\mathbf{r)}$ in terms of scalar $Y_{jm}$ and vector spherical harmonics$~%
\mathbf{Y}_{jm\pm },\mathbf{X}_{jm}$, 
\begin{align}
\mathcal{\psi }_{+}& =\psi _{+0}\mathbf{1}+\boldsymbol{\psi }_{+}\mathbf{%
\cdot }\boldsymbol{\sigma }  \notag \\
\psi _{+0}& =(u_{j0j}^{+}/r)Y_{jm};~\boldsymbol{\psi }%
_{+}=(u_{j+11j}^{+}(r)/r)\mathbf{Y}_{jm+}+(u_{j-11j}^{+}(r)/r)\mathbf{Y}%
_{jm-}+(u_{j+11j}^{+}(r)/r)\mathbf{X}_{jm-}.
\end{align}

Of crucial importance is that the 16 component or 4x4 matrix form of TBD
wave function must satisfy CM energy dependent norm: 
\begin{equation}
\int d^{3}x[\psi ^{\dag }(1+4w^{2}\beta _{1}\beta _{2}\partial \Delta
/\partial w^{2})\psi ]\equiv \int d^{3}x\psi ^{\dag }\mathcal{L}\psi =1,
\end{equation}%
with the matrix $\Delta (\mathbf{r)=}\gamma _{51}\gamma _{52}[L-\gamma
_{1}\cdot \gamma _{2}\mathcal{G}]/2$ displaying the core scalar and vector
interactions. In 4x4 matrix form we have 
\begin{equation}
\mathcal{\psi }\mathcal{=}\exp (F)[\cosh K\Psi (\mathbf{r)}-\sinh K%
\boldsymbol{\Sigma}\cdot \mathbf{\hat{r}}\Psi \mathbf{(\mathbf{r)}}%
\boldsymbol{\Sigma}\cdot \mathbf{\hat{r}]}\equiv \mathcal{K}\Psi (\mathbf{r)}%
,
\end{equation}%
giving us the norm in terms $\mathcal{\psi }$ and the Pauli reduction
solution $\Psi (\mathbf{r)}$. Thus in place of the Naive Norm (NN), 
\begin{equation*}
\int d^{3}xTr\Psi (\mathbf{r)}^{\dag }\Psi (\mathbf{r)}=1,
\end{equation*}%
we have the Two-Body Dirac Norm (TBDN), 
\begin{equation}
\int d^{3}xTr\mathcal{\psi }^{\dag }\mathcal{L\psi }=\int d^{3}xTr\left( 
\mathcal{K}\Psi (\mathbf{r)}\right) ^{\dag }\mathcal{LK}\Psi (\mathbf{r)}=1.
\end{equation}%
Using the (TBDN) norm has a significant effect on the decay amplitude and
rate compared to that of the (NN) norm.

For pseudoscalar meson decay, the triplet wave function is zero ($%
\boldsymbol{\psi }_{+}$ $=0$) $~$while for vector wave functions the singlet
wave function ($\psi _{+0}$ $=0$) is zero. The final step of the decay
formalism consists of substituting the resulting forms (\ref{frm}) into Eq.(%
\ref{amp}) for the matrix element and performing the numerical integration.

%TCIMACRO{\TeXButton{B}{\begin{table}[h] \centering}}%
%BeginExpansion
\begin{table}[h] \centering%
%EndExpansion
\begin{tabular}{llllllll}
& Expt. & TBDE(TBDN) & TBDE(NN) &  &  &  &  \\ 
$\pi^{0}(^{1}S_{0}-0.135)$ & 7.72$\pm$.04$~\mathrm{eV}$ & 8.73 \textrm{eV} & 
33.5 \textrm{eV} &  &  &  &  \\ 
$\eta_{c}(1^{1}S_{0}-2.976)$ & 7.4$\pm$1.0 \textrm{keV} & 6.20 \textrm{keV}
& 6.18 \textrm{keV} &  &  &  &  \\ 
$\eta_{c}(2^{1}S_{0}-3.263)$ & 1.3$\pm$0.6 \textrm{keV} & 3.36 \textrm{keV}
& 1.95 \textrm{keV} &  &  &  &  \\ 
$\chi_{0}(1^{3}P_{0}-3.415)$ & 2.6$\pm$0.65 \textrm{keV} & 3.96 \textrm{keV}
& 3.34 \textrm{keV} &  &  &  &  \\ 
$\chi_{2}(1^{3}P_{2}-3.556)$ & 0.53$\pm$0.09 \textrm{keV} & 0.743 \textrm{keV%
} & 0.435 \textrm{keV} &  &  &  & 
\end{tabular}
\caption{Meson 2 $\gamma$ Decay Rates\label{key}} 
%TCIMACRO{\TeXButton{E}{\end{table}}}%
%BeginExpansion
\end{table}%
%EndExpansion

The results are for the $\pi^{0}$ is very encouraging being off by only
13\%. \ This is in sharp contrast to the 3 order of magnitude errors in most
potential model applications. \ If we use the naive norm (NN) and amplitude
then we obtain 33.5 eV. The results for the $\eta_{c}$ meson is about the
same degree of accuracy. Work by Ackleh and Barnes \cite{ack} and earlier
ones by Haynes and Isgur \cite{hne} found it necessary to include results
that appealed to effective field theories for the decay instead of the
strictly microscopic approach we have taken here.

In summary, the \ constraint TBDE formalism gives a covariant two-body wave
equation that a) provides comprehensive account for entire meson spectrum,
b) is rigorously tested in QED, c) displays a remarkable connection between
pion and singlet positronium that accounts for Goldstone boson-like behavior
of pion, d) leads to a formalism that works well for light and heavy
two-photon meson decays.


\vspace*{-0.0cm}

\begin{thebibliography}{9}
\bibitem{bckr} H. W. Crater, R. Becker, C. Y. Wong and P. Van Alstine, Phys.
Rev. \textbf{D46}, 5117 (1992)

\bibitem{cr04} Details may be found in \ H. W. Crater and P. Van Alstine,
Phys. \textbf{D70, } 034026 (2004) and in references contained therein.

\bibitem{liu} B. Liu and H. W. Crater, Phys. Rev \textbf{C67,} 024001 (2003).

\bibitem{cr06} H. Crater, C. Y. Wong, and P. Van Alstine,Phys. Rev. \textbf{%
D74}, 054028 (2006)

\bibitem{szcr} H. Sazdjian, J. Math. Phys. \textbf{28} 2618 (1987), H. W.
Crater and P. Van Alstine, Ann. Phys. (N.Y.) \textbf{148} , 57 (1983).

\bibitem{adlr} S. L. Adler and T. Piran, Phys. Lett., \textbf{117B}, 91
(1982) and references contained therein.

\bibitem{isg} S. Godfrey and N. Isgur, Phys. Rev. D1 \textbf{32}, 189 (1985)

\bibitem{ack} E. S. Ackleh and T. Barnes, Phys. Rev. D \textbf{45,} 232,
(1992).

\bibitem{hne} C. Hayne and N. Isgur, Phys. Rev. D \textbf{25}, 1944, (1982).
\end{thebibliography}
\end{document}